\documentclass[aip,rsi,preprint,graphicx]{revtex4-1} 
\usepackage[utf8]{inputenc}
\usepackage[english]{babel}
\usepackage{graphicx}

\begin{document}

\title{Definition of design guidelines, construction and performance of an ultra-stable scanning tunneling microscope for spectroscopic imaging} 

\author{Irene Battisti}
\affiliation{Leiden Institute of Physics, Leiden University, 2333 CA Leiden, The Netherlands}

\author{Gijsbert Verdoes}
\affiliation{Leiden Institute of Physics, Leiden University, 2333 CA Leiden, The Netherlands}
\affiliation{Fine Mechanical Department, Leiden University, 2333 CA Leiden, The Netherlands}

\author{Kees van Oosten}
\affiliation{Leiden Institute of Physics, Leiden University, 2333 CA Leiden, The Netherlands}
\affiliation{Fine Mechanical Department, Leiden University, 2333 CA Leiden, The Netherlands}

\author{Koen M. Bastiaans}
\affiliation{Leiden Institute of Physics, Leiden University, 2333 CA Leiden, The Netherlands}

\author{Milan P. Allan}
\email[]{allan@physics.leidenuniv.nl}
\affiliation{Leiden Institute of Physics, Leiden University, 2333 CA Leiden, The Netherlands}

\date{\today}

\begin{abstract}
Spectroscopic-imaging scanning tunneling microscopy is a powerful technique to study quantum materials, with the ability to provide information about the local electronic structure with subatomic resolution. However, as most spectroscopic measurements are conducted without feedback to the tip, it is extremely sensitive to vibrations coming from the environment. This requires the use of laboratories with low-vibration facilities combined with a very rigid microscope construction.
In this article, we report on the design and fabrication of an ultra-stable STM for spectroscopic-imaging measurements that operates in ultra high vacuum and at low temperatures (4\,K). We perform finite element analysis calculations for the main components of the microscope in order to guide design choices towards higher stiffness and we choose sapphire as the main material of the STM head. By combining these two strategies, we construct a STM head with measured lowest resonant frequencies above $f_0 =13$\,kHz for the coarse approach mechanism, a value three times higher than previously reported, and in good agreement with the calculations. With this, we achieve an average vibration level of $\sim 6$\,fm$/\sqrt{\textrm{Hz}}$, without a dedicated low-vibration lab. We demonstrate the microscope's performance with topographic and spectroscopic measurements on the correlated metal Sr$_2$RhO$_4$, showing the quasiparticle interference pattern in real and reciprocal space with high signal-to-noise ratio. 
\end{abstract}


\maketitle 

\section{Introduction}

Spectroscopic-imaging scanning tunneling microscopy (SI-STM) is an important tool to study the properties of quantum materials\cite{Series2012,Lee2005,Yazdani2016a}. The power of the technique stems from the ability of performing tunneling spectroscopy experiments with high energy and spatial resolution, allowing one to map the electronic local density of states (LDOS) onto a simultaneously acquired image of the atomic lattice.  
By observing quasiparticle interference (QPI), one can also access the reciprocal space, providing information about the band structure below and above the Fermi level\cite{Petersen1998, Crommie1993,Hoffman2002}.
Since its application to the field of correlated electron systems, and in particular to high-T$_{\rm{c}}$ superconductors, SI-STM allowed, among others, to image the impact of individual dopant atoms on the electronic structure\cite{Pan2000, Hudson2001}, to visualize nanoscale electronic disorder\cite{Lang2002, McElroy2005}, to image local symmetry breaking \cite{Kohsaka2007, Lawler2010c} and to determine the gap structure by QPI imaging\cite{Hoffman2002, Lee2012}. 

In order to achieve the data quality that is needed to understand the properties of correlated electron systems, there are some strict requirements for the design of the microscope. In a typical STM experiment, the tip scans over the sample to acquire a topographic image using a feedback mechanism that keeps the tunneling current constant by controlling the tip-sample distance. The feedback thus partially corrects for vibrations transmitted to the tunnel junction, which are usually the main source of noise in STM. When performing a spectroscopy experiment, instead, the feedback loop is switched off to measure a dI/dV spectrum, which is proportional to the LDOS. During this spectroscopic measurement the instrument is very sensitive to external vibrations: Given the exponential dependence of the tunneling current on the tip-sample distance, even the smallest vibrations will make the tunneling current, and thus the dI/dV spectrum, significantly noisier. In a typical SI-STM measurement, $10^5 - 10^6$ of such spectra distributed on a fine spacial grid are consecutively acquired over several days. This demands extreme stability of the microscope in order to achieve the signal-to-noise ratio (S/N) needed to study the properties of quantum materials.

Traditionally, high stability is achieved by combining the use of a vibration isolation table with a rigid microscope construction. Both are essential to obtain good data quality. However, during the design phase, it is mostly important to focus on the rigidity of the STM head, and thus of the tip-sample junction, because it is the factor that can drastically improve the performance of the microscope. This is the focus of the present paper.

A good way of quantifying the rigidity of the STM head is via its lowest resonant frequency $f_0$, usually in the kHz range:  vibrations with frequencies $f \ll f_0$ will not be transmitted to the junction, because tip and sample will move in phase when excited with such frequencies, and only vibrations with  $f \geq f_0$ will enhance the noise.  A good criterion to achieve a high S/N is thus to build a STM head with the highest possible value of $f_0$\cite{Chen2000, White2011}.

\begin{figure*}
\centering
\includegraphics[width=15cm]{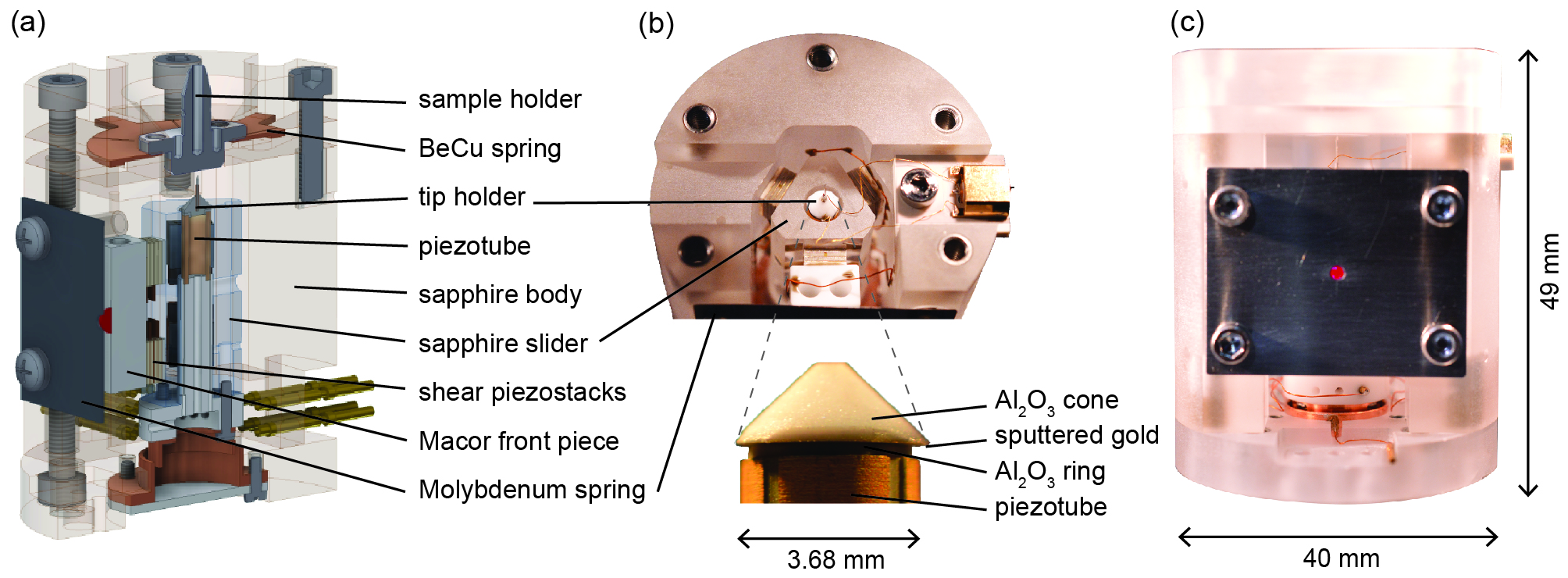}
\caption{(a) Cross section of the 3D CAD rendering of the STM head, with the main components being pointed out. (b) Top view of the open STM head, where the slider construction and the tip holder can be seen. In the detail, picture of the tip holder during the construction phase. The conical Al$_2$O$_3$ holder has been sputtered with gold on the bottom surface to provide a ground shield for the current signal (measured from the tip). In between the latter and the piezotube, a small Al$_2$O$_3$ ring is glued. (c) Picture of the fully assembled STM head. }
\label{fig:STM}
\end{figure*}

In the following, we report on the design, construction and performance of a home-built STM for spectroscopic imaging measurements, defining guidelines in order to obtain a very stable microscope. The instrument operates in ultra high vacuum (base pressure $p =3\times 10^{-10}$\,mbar ) and at low temperatures ($T=4$\,K).
In Section \ref{sec:design}, we focus on the STM head design and construction, and we explain how we achieve very high stiffness via both material choice and the use of finite element analysis (FEA) calculations to improve the geometry. In Section \ref{sec:system}, we give an overview of the cryogenic insert and of the vibration isolation table. Finally, in Section \ref{sec:perf} we show the performance of the STM, and report vibrational noise spectra and high quality topographic and spectroscopic data measured at 4\,K on the correlated metal Sr$_2$RhO$_4$.

\section{\label{sec:design}Design an ultra-stiff STM head using finite element analysis}

In order to obtain a very stiff cryogenic-compatible STM head, we start from the design of Pan et al. \cite{Pan1999} and optimize both materials and geometry in order to maximize the lowest resonant frequency $f_0$. 

Regarding the materials, we choose to use oriented single-crystalline sapphire for the body of the STM head. As shown from White et al. \cite{White2011}, sapphire is the ideal material for the main parts of the STM head because it is one of the stiffest insulating materials available and it has an excellent low temperature thermal conductivity. Recently, sapphire has become easier and more affordable to machine, allowing us to build on the approach of White and al.  \cite{White2011} and design more complicated and detailed sapphire parts for the STM head.

Here, we mainly focus on the optimization of the geometry, which we achieve by finite element analysis (FEA) calculations performed with the commercially available software Comsol \cite{COMSOL}. FEA calculations have been used before to improve the geometry of various scanning probe techniques \cite{Kindt2004,Ast2008,Maess2008,Braunsmann2010}, and in particular of high-scanning speed STM \cite{Rost2005,Rost2009a,Yanson2013}. Here, we use them to improve the geometry of a low-temperature STM for spectroscopic imaging. 

In Fig.\,\ref{fig:STM}, a CAD rendering and pictures of the assembled STM head are shown.
The head consists of a hollow cylindrical sapphire body open on one side, inside which a triangular polished sapphire prism (the slider) is clamped via six shear piezo stacks (PI Ceramic) and a molybdenum spring. This main construction provides the coarse approach mechanism (the walker), with the slider moving reliably up and down at all temperatures when high voltage pulses are applied to the shear piezo stacks. The piezotube (EBL Products, Inc.), providing the scanning motion both in the Z (vertical) and in the XY (in-plane) directions, is glued with non-conductive epoxy (Epotek H74F) on top of a Macor support which is in turn fixed inside the slider. It has 4 outer electrical contacts for deflections in the $\pm$X, $\pm$Y directions and one inner contact for contractions and elongation in Z. The tip is mounted on a tip holder glued on top of the piezotube. The tip holder assembly is optimized for low weight, in order to increase the resonant frequencies of the scanner, and for low capacitance, in order to allow for RF-STM measurements \cite{Bastiaans2018a}. It consists of a small Al$_2$O$_3$ ring glued directly on the piezotube, and a conical Al$_2$O$_3$ holder glued on the ring, as shown in Fig.\,\ref{fig:STM}(b). This last piece hosts a small molybdenum cylinder inside which the tip can be clamped. The bottom surface of the conical piece is additionally sputtered with 450\,nm of gold to provide a ground shield for the tip wire. With this assembly we obtain a tip-to-sample capacitance of 180\,fF (at tunneling distance, with tip retracted) and a tip-to-ground shield capacitance of 2\,pF. The tip-to-ground shield capacitance should be as low as possible to facilitate RF-STM measurements, and the value we measure is reduced of more than a factor 10  with respect to commercially available instruments. 

\begin{figure}
\centering
\includegraphics[width=8.2cm]{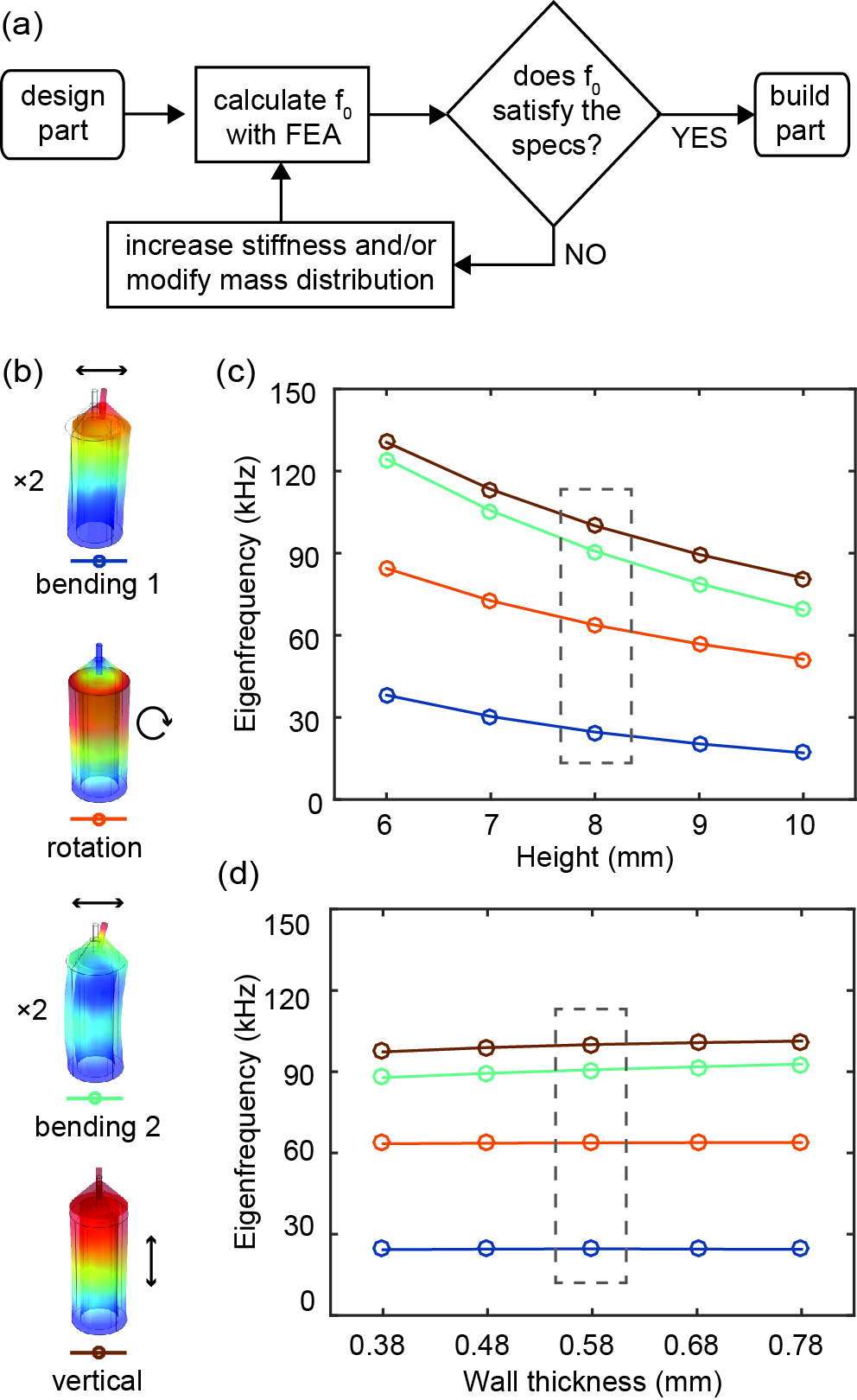}
\caption{(a) Flow chart explaining the procedure applied to several components of the STM head in order to achieve higher resonant frequencies. (b) Mode shapes of the first six mechanical resonances of the piezotube as computed with Comsol, as a legend for panels (c-d). The modes are shown as (exaggerated) displacements with colors from red (max displacement) to blue (no displacement).   
(c) Variation of eigenfrequencies of the piezotube's modes with respect to the piezotube height, with base diameter fixed. (d) Variation of eigenfrequencies of the piezotube's modes with respect to the wall thickness (with fixed outer diameter $D=3.68$\,mm). The dashed boxes indicate the eigenfrequencies for the final chosen dimensions. }
\label{fig:PZT}
\end{figure}

\subsection{\label{subsec:designFEA}Design principles from finite element analysis}
The optimization procedure that we applied to several components of the STM head in order to achieve high resonant frequencies is shown in Fig.\,\ref{fig:PZT}(a). As an example, we show here how we apply this procedure to the piezotube. After importing the initial design in Comsol, we compute the mechanical eigenfrequencies of the first vibrational modes with the base of the piezotube fixed, as shown in Fig.\,\ref{fig:PZT}(b). We can identify two double degenerate bending modes, a rotational mode and a vertical mode. 
We then progressively change the dimensions of the part and recompute the eigenfrequencies, in order to understand which dimensions can be optimized to increase the stability. In the case of a simple tube, this problem could also be solved analytically, however here we want to show the procedure that than we will apply to more complicated geometries. We find that is beneficial to lower the height [Fig.\,\ref{fig:PZT}(c)], while changing the wall thickness (with fixed outer diameter) does not have a significant effect [Fig.\,\ref{fig:PZT}(d)]. In general, the optimization towards higher resonant frequencies can be limited by functional requirements and geometrical constraints. For the piezotube, the final design is chosen as a compromise between stability and desired XY scanning range of $\geq$\,350\,nm at 4\,K.

A similar analysis is performed on other STM head components, including the sapphire slider, the STM body, the Macor front piece and the sample holder. For instance, the simulations indicate that to reach a higher $f_0$, the STM head will benefit from a lower aspect ratio, a shorter and lighter sapphire slider and a lighter Macor front piece. 

\begin{figure}
\centering
\includegraphics[width=8.2cm]{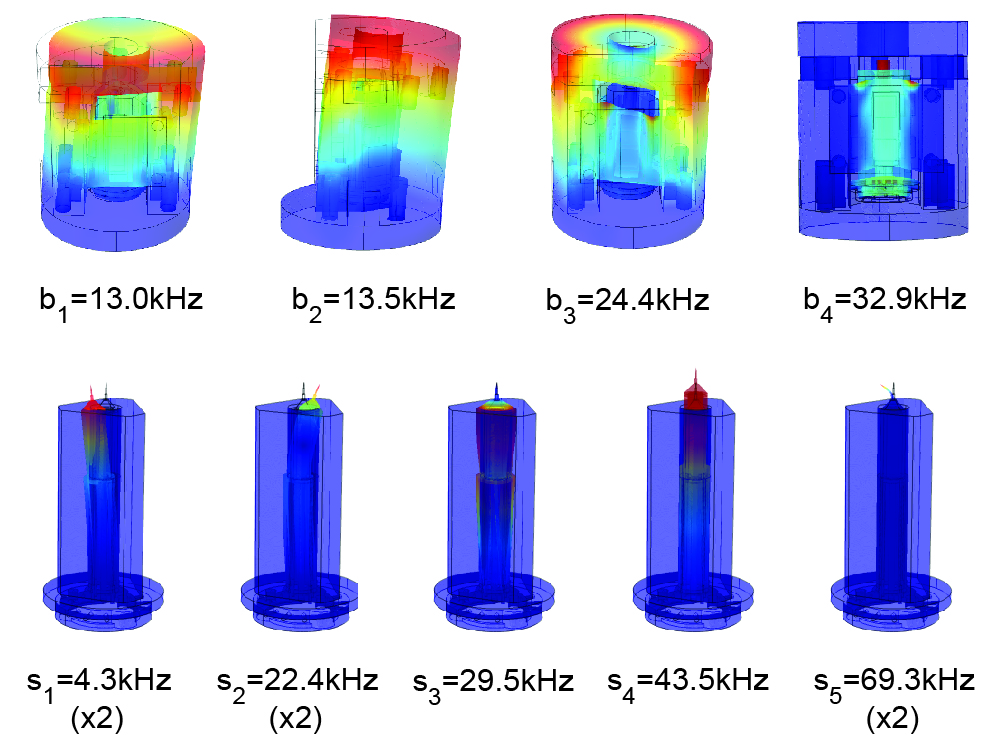}
\caption{Main vibrational modes of the STM head, as calculated with Comsol, with the base of the STM fixed. All the modes shown are calculated for the full design, but to ease visualization the modes that belong to the scanner assembly ($s_1$-$s_5$) are shown separately. See the main text, Sec.\,\ref{subsec:designFEA}, for a description of the modes.} 
\label{fig:head}
\end{figure}

Finally, we also performed FEA calculations on the whole assembly of the STM head, simplified by removing small details in the design, to allow for a simpler mesh. This can provide further insights about the design choices, and moreover it can be directly compared to the measured values of the eigenfrequencies. The most significant modes that we obtain from FEA calculations of the full STM head are shown in Fig.\,\ref{fig:head}. They can be divided in two groups: the modes belonging to the body of the STM ($b$-modes in Fig.\,\ref{fig:head}) and the modes belonging to the scanner assembly ($s$-modes). The modes $b_1$, $b_2$ are almost degenerate and correspond to the bending modes of the full head, $b_3$ corresponds to the rotational mode and $b_4$ to the motion of the slider inside the head. The modes $s_1$ and $s_2$ are double degenerate and correspond to first order and second order bending mode of the scanner, $s_3$ is the rotational mode of the scanner, $s_4$ the vertical mode of the scanner and $s_5$ is the bending mode of the tip.

Intuitively, one could think that the walker assembly used for coarse approach is the weakest point of the STM head design, and therefore that the lowest vibrational mode should correspond to the vertical motion of the slider inside the scanner ($b_4$). In our simulations, however, the main body vibrational modes and the scanner bending modes have lower frequencies, showing that the limiting factor for $f_0$ is mainly the STM head geometry. Clearly, the calculated value of the mode $b_4$ depends strongly on the stiffness of the shear piezo stacks that we feed into the simulation. The value of $b_4$ shown in Fig.\,\ref{fig:head} is obtained with half the stiffness reported in the specifications, to take into account the effect of glue and other factors involved in the mounting. Furthermore, the measured resonant frequencies reported in section \ref{sec:meas_f0} confirm that the slider construction is not the weakest point of the design.

\subsection{Measurement of resonant frequencies}
\label{sec:meas_f0}
\begin{figure}
\centering
\includegraphics[width=7cm]{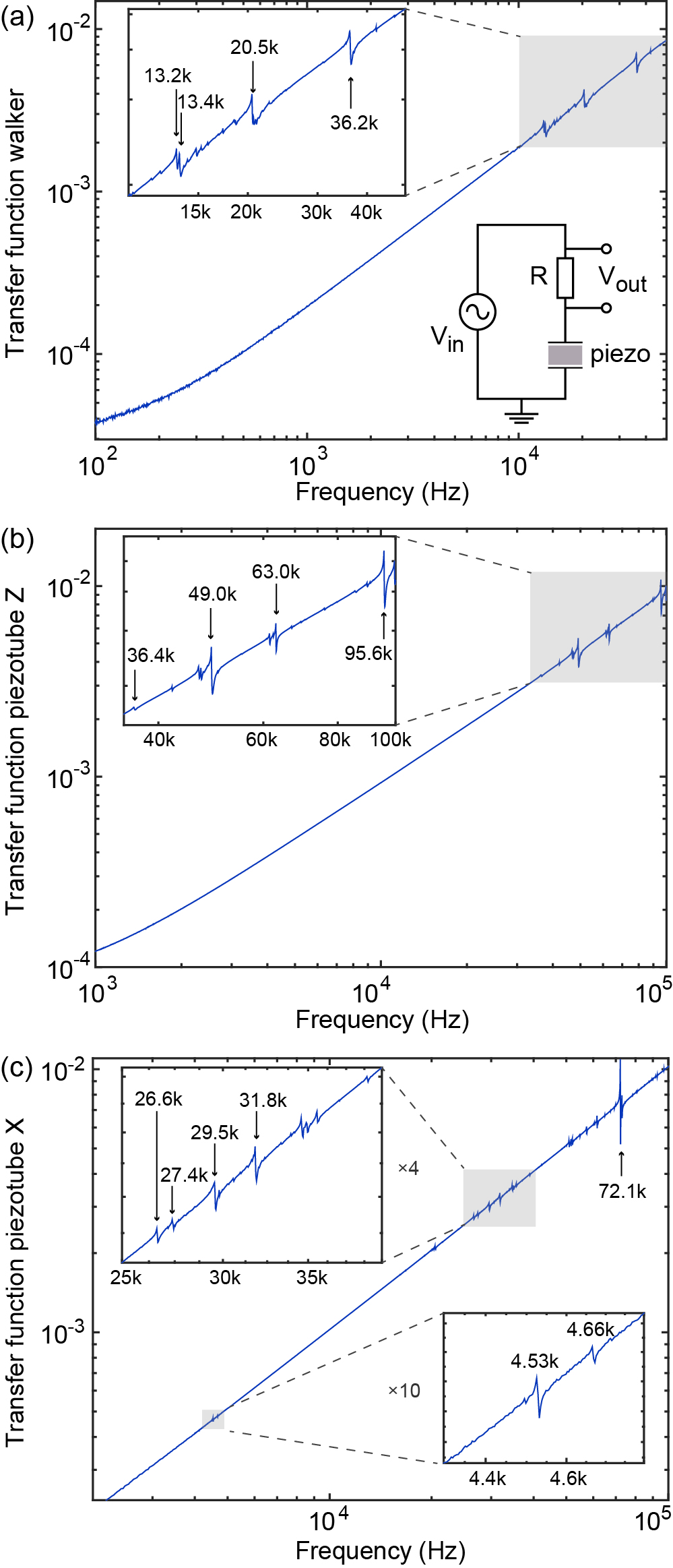}
\caption{Measured transfer functions $V_{\rm{out}}/V_{\rm{in}}$ of the coarse approach system and of the piezotube. (a) Transfer function of the shear piezo stacks coarse approach mechanism. A scheme of the electronic circuit is shown at the bottom right corner, where $R=11$\, $\Omega$. In the inset, a zoom of the area shaded in gray is shown. (b) Transfer function of the piezotube vertical excitation. In the inset, a zoom of the gray area is shown. (c) Transfer function of the piezotube bending excitation. In the insets, zooms of the gray areas are shown (note the different scales).} 
\label{fig:tr}
\end{figure}

After the construction of the STM head, we measure both the eigenfrequencies of the walker assembly and of the piezotube assembly independently and we compare them with the calculations. The measurement is performed \textit{in situ} with the head mounted in the cryogenic insert at room temperature and under vacuum. A lock-in amplifier is used to excite the piezo elements and to read out their response. The applied excitation is a sinusoidal signal of amplitude 1\,$V_{\rm{peak-peak}}$ and sweeping frequency. A scheme of the circuit is presented in the inset of Fig.\,\ref{fig:tr}(a)\cite{Rost2005}.
Thanks to the intrinsic properties of piezoelectric materials, the electrical excitations $V_{\rm{in}}$ provided by the lock-in is converted into a mechanical excitation. When the excitation frequency corresponds to one of the STM eigenfrequencies, its amplitude gets enhanced, resulting in an enhanced electrical signal from the piezos $V_{\rm{out}}$ that is detected by the lock-in \cite{Rost2005,Rost2009a}.

Figure\,\ref{fig:tr}(a) shows the transfer function obtained when exciting the six shear piezo stacks for the coarse approach connected in parallel. The main result is that we do not observe any resonant mode of the coarse approach assembly before 10\,kHz, with the first strong peaks appearing at $\sim$13\,kHz. We also performed the measurement at 4\,K (not reported here), leading to slightly higher resonance frequencies due to the different stiffness of the materials at low temperatures and to thermal contractions.
Remarkably, the value we achieve is a factor three better than the previously reported eigenfrequencies for a sapphire STM head \cite{White2011}, where the first strong peak in the coarse approach excitation appears at 4.6\,kHz. 

Additionally, we can now compare the measured resonances with the computed ones. We can easily identify the first two almost degenerate peaks at 13.2\,kHz and 13.4\,kHz as the two bending modes of the STM body ($b_1, b_2$). The peak at 20.5\,kHz is more difficult to identify, but it could correspond to the rotational mode $b_3$. We interpret the peak measured at 36.2\,kHz as the mode $b_4$, given that in the FEA calculations $b_4$ is the only mode we find in the frequency window between 30\,kHz and 50\,kHz, and in the measurement we do not observe any other peak before $\sim$80\,kHz. 

For the piezotube, we can measure the transfer function while exciting either the vertical modes (which are the most crucial ones) or the bending modes (which have less influence on the noise), simply by applying the excitation to different electrodes.
In Fig.\,\ref{fig:tr}(b) we show the transfer function of the vertical modes, measured with the outer electrical contacts ($\pm$X,$\pm$Y) connected in parallel against the inner contact (Z) acting as ground. The small peak at 36.4\,kHz results from coupling to the mode $b_4$, and the peak at 49.5\,kHz can be identified as the vertical mode $s_4$ of the full scanner assembly. The peaks at 63.0\,kHz and at 95.6\,kHz are more difficult to identify, but could correspond to the rotational and vertical mode of the piezotube alone, that the calculations give respectively at 63.7\,kHz and 100.0\,kHz (from Fig.\,\ref{fig:PZT}).
In Fig.\,\ref{fig:tr}(c) we show the transfer function of the bending modes, measured between +X and -X contacts. The two very small peaks measured at 4.53\,kHz and 4.66\,kHz correspond to the first order bending modes of the piezotube $s_1$. The peaks at 26.6\,kHz and 27.4\,kHz could correspond to the second order bending modes $s_2$, and the peak at 29.5\,kHz to the rotational mode $s_3$. We do not find correspondence to the calculations for the other small peaks, and the strong peak at 72.1\,kHz could correspond to the bending mode of the tip $s_5$, even if we are doubtful that this could lead to such a sizable response of the piezotube.

The lowest resonant frequencies of the STM head thus correspond to the bending modes of the piezotube and of the STM head, which are substantially less critical for STM measurements than the vertical modes. The walker assembly, often a critical point for the vertical stability, proves to be the stiffest reported to date for a Pan-style head. Taken together, this confirms the exceptional stiffness of our STM head.

\section{\label{sec:system}Design of cryogenic insert and vibration isolation table}

\begin{figure}
\centering
\includegraphics[width=8.2cm]{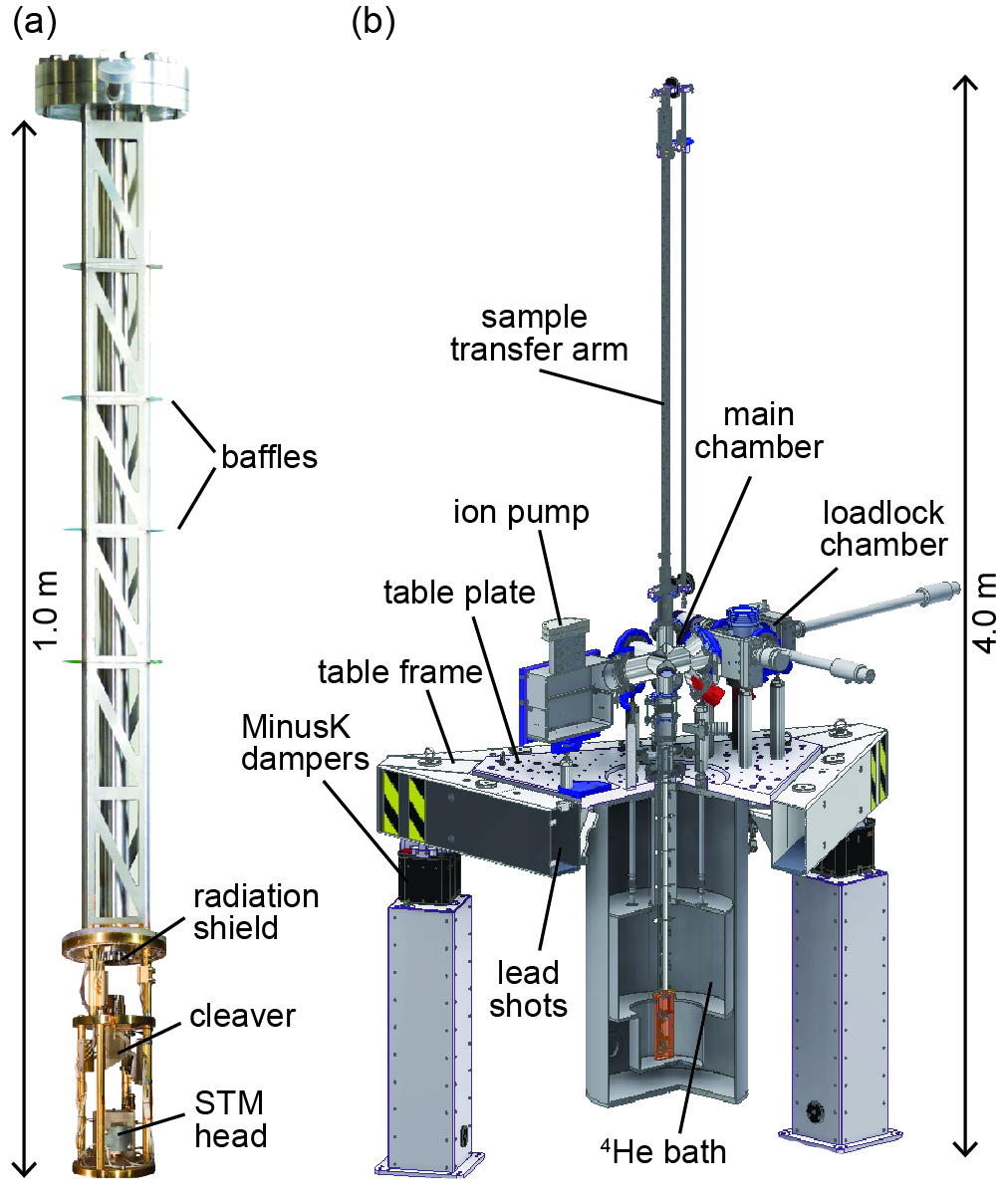}
\caption{(a) Picture of the cryogenic insert. (b) Three quarters section of the CAD rendering of the full system.}
\label{fig:tot}
\end{figure}

The STM head is mounted at the bottom of a home-built cryogenic insert placed in a $^{4}$He bath cryostat and it operates at the base temperature of 4.3\,K.
The insert is rigidly connected to the UHV chamber and the table, which is in turn suspended on negative-stiffness vibration isolators (Minus K\textsuperscript{\textregistered}). A picture of the cryogenic insert and a CAD rendering of the full system are shown in Fig.\,\ref{fig:tot}.

\subsection{Cryogenic insert}
The bottom part of the insert, where the STM is rigidly connected without the use of springs, is made of gold-plated oxygen-free copper to optimize the thermal connection to the $^{4}$He bath. It is provided with a radiation shield and a cold cleaving stage that can respectively be placed and actuated by the sample transfer mechanism. The connection to the room temperature UHV chamber is made of thin-walled stainless steel tubes, that are used for pumping and wiring.

The design of the cryogenic insert has also been optimized for maximum stability. For this reason, we opted for a reinforced insert structure that features four stainless steel tubes. The central tube has a outer diameter of 31.0\,mm and wall thickness of 0.5\,mm and it is used as sample transfer line and main pumping line. The three smaller tubes (of outer diameter 12.7\,mm and wall thickness 0.25\,mm) are used for wiring and are placed in a triangular pattern around the central one. We find that choosing three tubes for wiring, instead of two, drastically improves the stability for the pendulum mode, with the resonant frequency corresponding to this mode increasing by 50\,$\%$. Furthermore, we provide additional stability by connecting the tubes with thin (0.3\,mm) stainless steel plates. These plates have triangular holes in order to reduce the heat conductivity, and are laser spot welded directly to the tubes. Around the tubes, and perpendicular to them, we place four stainless steel baffles that act as radiation shields inside the dewar neck, to reduce the radiation heat transfer to the $^{4}$He bath. 

All the extra features that we added to improve the stability imply a higher heat conduction to the $^{4}$He bath, and a compromise has to be found between stability and desired helium consumption. With our particular design we still obtain an excellent helium consumption of 0.18\,L/h (4.3\,L/day), which ensures a very quiet helium boil-off noise and allows for $>$7 days long measurements.

\subsection{Vibration isolation table}
The vibration isolation table is made of two parts: a rigid stainless steel hollow frame with triangular shape and a 40-mm-thick triangular aluminum plate. The plate is rigidly connected to the frame by screws, and the assembly is designed with the specifications of not having resonant frequencies below 300\,Hz. Both the UHV chamber and the dewar are rigidly fixed to the aluminum plate. The frame is mounted on top of three negative-stiffness vibration isolators (Minus K\textsuperscript{\textregistered} 800CM-1) with resonant frequency 0.5\,Hz. In order to dampen acoustic-induced noise and vibrations and to increase the mass of the table, the frame is filled with lead shots to reach a weight of the total system of $\sim$\,1000\,kg. 

The full system is standing on a measurement island that is decoupled from the surrounding walking floor, but is not additionally isolated with dampers from the foundations.

\section{\label{sec:perf}Performance}

\subsection{Vibrations and I/z noise at the tunneling junction}
\begin{figure}
\centering
\includegraphics[width=8.2cm]{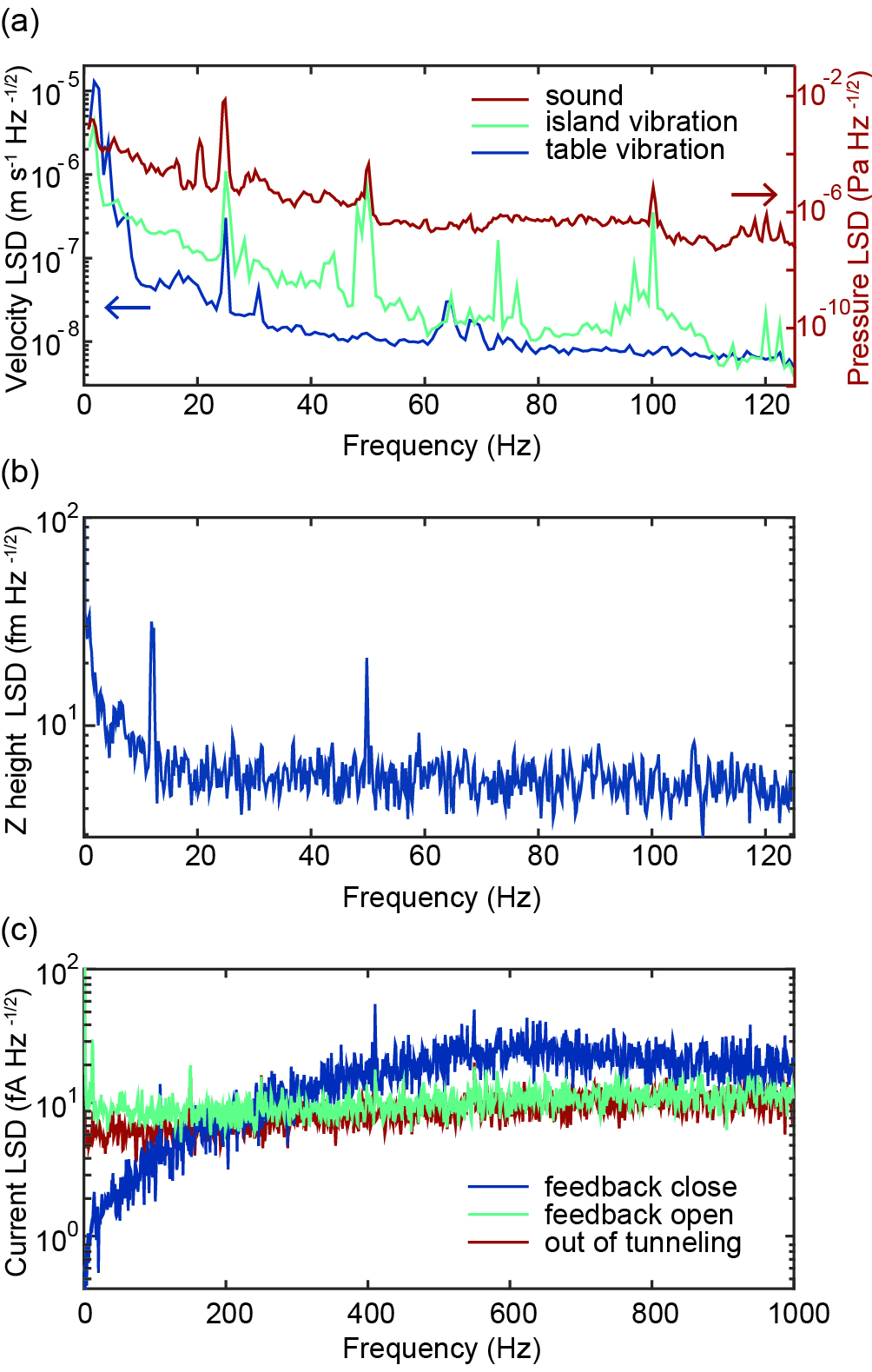}
\caption{ (a) Linear spectral density (LSD) of the velocity measured in the Z direction on the island and on the STM table with a seismometer (Guralp CMG-40T). In the same graph, with scale on the right axis, acoustic noise is reported as LSD of the pressure difference measured with a low-frequency microphone (G.R.A.S. 46AF). (b) LSD of the Z height noise at the tunnel junction measured on (Pb,Bi)$_2$Sr$_2$CuO$_{6+x}$ at setup condition V\,=\,300\,mV and I\,=\,150\,pA.  (c) LSD of the tunneling current at same tunnel junction as (b), measured out of tunneling as well as in tunneling with feedback loop closed and open. }
\label{fig:vib}
\end{figure}

As a first test of the performance of our microscope, we measure the vibrational noise level at the tunneling junction. 

We first characterize the noise level of our laboratory by measuring both vibrations with a seismometer (Guralp CMG-40T) and acoustic noise with a low-frequency microphone (G.R.A.S. 46AF). The measurement is reported as linear spectral density (LSD) of the velocity for the vibrations and of the pressure difference for the acoustics. The spectra are obtained by Fourier transforming one minute of real time data with the Welch method using a Hamming window. 

We report in Fig.\,\ref{fig:vib}(a) typical spectra of (i) the vibrations measured on the island where the table is standing, (ii) the vibrations measured on the table, (iii) the sound measured in the surrounding room. It can immediately be noticed that the table dampens the general vibrational noise of about one order of magnitude after 10 Hz, and it reduces many of the peaks. The table is less effective in reducing the sound-induced noise, as it can be noticed from the peak at 24 Hz that is only partially dampened. 

In Fig.\,\ref{fig:vib}(b) we show the vibrational noise at the tunnel junction as the linear spectral density of the Z height (corresponding to the tip-sample distance) measured in feedback on a BSCCO sample ((Pb,Bi)$_2$Sr$_2$CuO$_{6+x}$) at setup condition V\,=\,300\,mV and I\,=\,150\,pA. Remarkably, the noise is very low, with an average vibration level of $\sim$\,6\,fm$/\sqrt{\rm{Hz}}$ and only two peaks with amplitude bigger than 20\,fm$/\sqrt{\rm{Hz}}$, respectively at 12\,Hz and 50\,Hz. 

In Fig.\,\ref{fig:vib}(c) we further show the linear spectral density of the current noise measured with the same tunneling junction and setup condition. With tip retracted, we achieve an average noise level of $\sim$8\,fA$/\sqrt{\rm{Hz}}$. The preamplifier we use (FEMTO\,LCA-4K-1G) has a noise floor of 6.5\,fA$/\sqrt{\rm{Hz}}$. 

In tunneling and with feedback loop closed we can see an increase in the noise to $\sim$20\,fA/$\sqrt{\rm{Hz}}$ above 200\,Hz, with the addition of some peaks reaching up to $\sim$50\,fA$/\sqrt{\rm{Hz}}$. The feedback takes care of reducing the low frequency noise. 

An important figure of merit for spectroscopic measurements is the out-of-feedback current noise (in tunneling and with feedback loop open). We obtain a noise level of $\sim$10\,fA$/\sqrt{\rm{Hz}}$, with only the peaks at 12\,Hz and 50\,Hz higher than 20\,fA$/\sqrt{\rm{Hz}}$, which ensures the capability of performing high quality spectroscopy experiments.

\subsection{Topograph and QPI on Sr$_2$RhO$_4$}

\begin{figure}
\centering
\includegraphics[width=8.2cm]{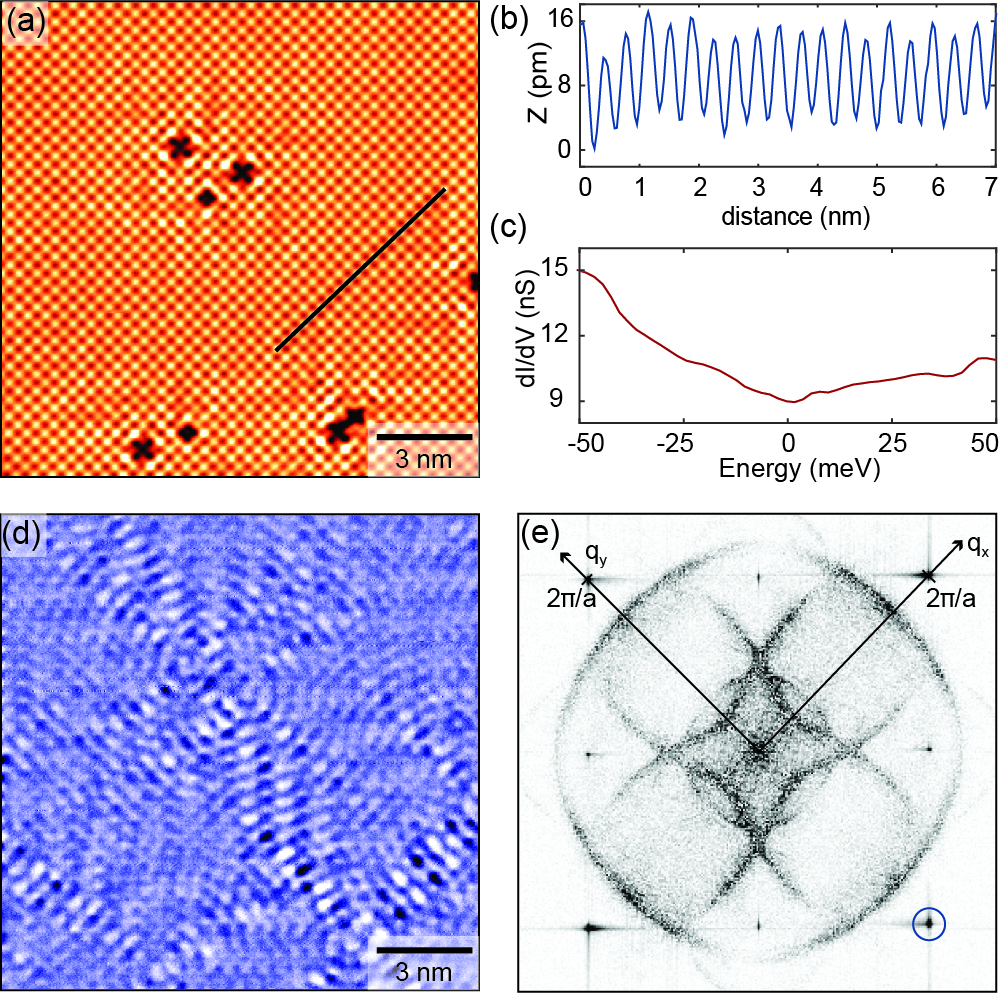}
\caption{STM measurements on Sr$_2$RhO$_4$. (a) Topograph in a field of view of $15\times15$\,nm$^2$. Setup condition V\,=\,-20\,mV, I\,=\,600\,pA. (b) Atomic corrugation profile along the black line in (a). (c) Example of a single dI/dV spectrum measured during a spectroscopic map. (d) Density of states measured simultaneously to the topograph in panel (a), showing the quasiparticle interference pattern in real space at energy $E=-20$\,meV. (e) Non-symmetrized Fourier transform of the conductance layer at the Fermi level acquired in a spectroscopic map over a $55\times55$\,nm$^2$ field of view. One of the Bragg peaks is highlighted in the blue circle.}
\label{fig:STMdata}
\end{figure}

To demonstrate the topographic and spectroscopic capabilities of the microscope, we show quasiparticle interference measurements on the correlated metal Sr$_2$RhO$_4$, that is known to have a highly two-dimensional electronic structure\cite{Perry2006, Baumberger2006, Craig2012a}.

The sample is cleaved at the 4\,K cleaving stage and immediately transferred to the STM head. 
Atomic resolution can be easily achieved, as shown in the $15\times15$\,nm$^2$ topograph in Fig.\,\ref{fig:STMdata}(a) and in the linecut along the black line plotted in Fig.\,\ref{fig:STMdata}(b). 
The cleaved surface shows the SrO layer, with the distance between the Sr atoms being 3.9\,\AA, and two different defects\cite{Stoger2014}. 
In Fig.\,\ref{fig:STMdata}(c) we plot a simple dI/dV spectrum acquired during a spectroscopic map. 
In Fig.\,\ref{fig:STMdata}(d) we show the real space imaging of QPI in the conductance layer at -20\, meV (with respect to the Fermi level), acquired simultaneously to the topograph in panel (a). The defects obviously act as scattering centers, creating an interference pattern that enhances the quantum mechanical wave nature of the quasiparticles. In Fig.\,\ref{fig:STMdata}(e) we show the Fourier transform of the real-space QPI pattern\cite{Craig2012a}, measured at the Fermi level when taking a spectroscopic map on a field of view of $55\times55$\,nm$^2$. The Fourier-transformed data is not symmetrized nor processed in any other way, and still shows sharp features with very high signal-to-noise ratio. 

A more comprehensive analysis, including a quantitative comparison with ARPES and quantum oscillation data\cite{Perry2006, Baumberger2006}, will be published elsewhere \footnote{in preparation}.

\section{Conclusions}
In this paper, we report on improved design principles, construction and performance of an ultra-stable STM for spectroscopic imaging. We show that by combining smart material choice and design improvement guided by FEA calculations we are able to achieve a very stiff STM head that guarantees the ability of obtaining outperforming data quality even in a lab not dedicated to low vibrations. We demonstrate the extreme stability of our instrument by obtaining an out-of-feedback current noise level of $\sim$10\,fA$/\sqrt{\rm{Hz}}$. Moreover, our data on the correlated metal Sr$_2$RhO$_4$ proves an excellent performance of the microscope in the acquisition of spectroscopic measurements.

\begin{acknowledgments}
We thank M. Rost, F. Galli, K. Heeck, M. Leeuwenhoek, T.-M. Chuang, P. Wahl, M. Hamidian, I. Firmo and J.C. Davis for valuable discussions, T. van Klingeren, M. Witlox, E. Wiegers, D. Oosterlee, A. Geluk, R. Koehler, L. Crama, and W. van der Geest for technical support, and A. Tamai, F. Baumberger and R. Perry for suggesting STM investigations on the rhodate samples and providing the crystals and unpublished data.

This work was supported by the Netherlands Organization for Scientific Research (NWO/OCW) as part of the Frontiers of Nanoscience (NanoFront) program and the VIDI talent scheme (Project No. 680-47-536).

\end{acknowledgments}

%

\end{document}